\documentclass[
    ,final            
  ]
  {aipproc}
\layoutstyle{6x9}
\usepackage[percent]{overpic}
\usepackage{graphicx}
\begin{document}
\title{\bf Measurement of the polarization observables $I^{s}$ and $I^{c}$ in the reaction 
$\boldmath \vec{\gamma}{\mathrm p}\rightarrow {\mathrm p}  \pi^0 \pi^0$ with the CBELSA/TAPS experiment}

\classification{\texttt{13.60.Le, 13.60.Rj, 13.88.+e}}

\author{V. Sokhoyan \\ for the CBELSA/TAPS Collaboration}{
  address={Helmholtz-Institut f\"ur Strahlen- und Kernphysik der Universit\"at Bonn, Germany}
}

%

\begin{abstract}

To unambiguously identify baryon resonances the measurement of polarization observables is of great importance.
With the CBELSA/TAPS experiment, located at the ELSA accelerator facility in Bonn, 
polarization observables have been determined. Using a linearly polarized photon beam impinging 
on a liquid hydrogen target, the polarization observables $I^{s}$ and $I^{c}$ for the reaction
$\vec{\gamma}\mathrm{p}\rightarrow\mathrm{p}\pi^{0}\pi^{0}$  have been determined for the first time.
\end{abstract}
\maketitle


\section{Introduction}

\begin{figure}[hbtp]
      \begin{overpic}[width= 8 cm,
]
{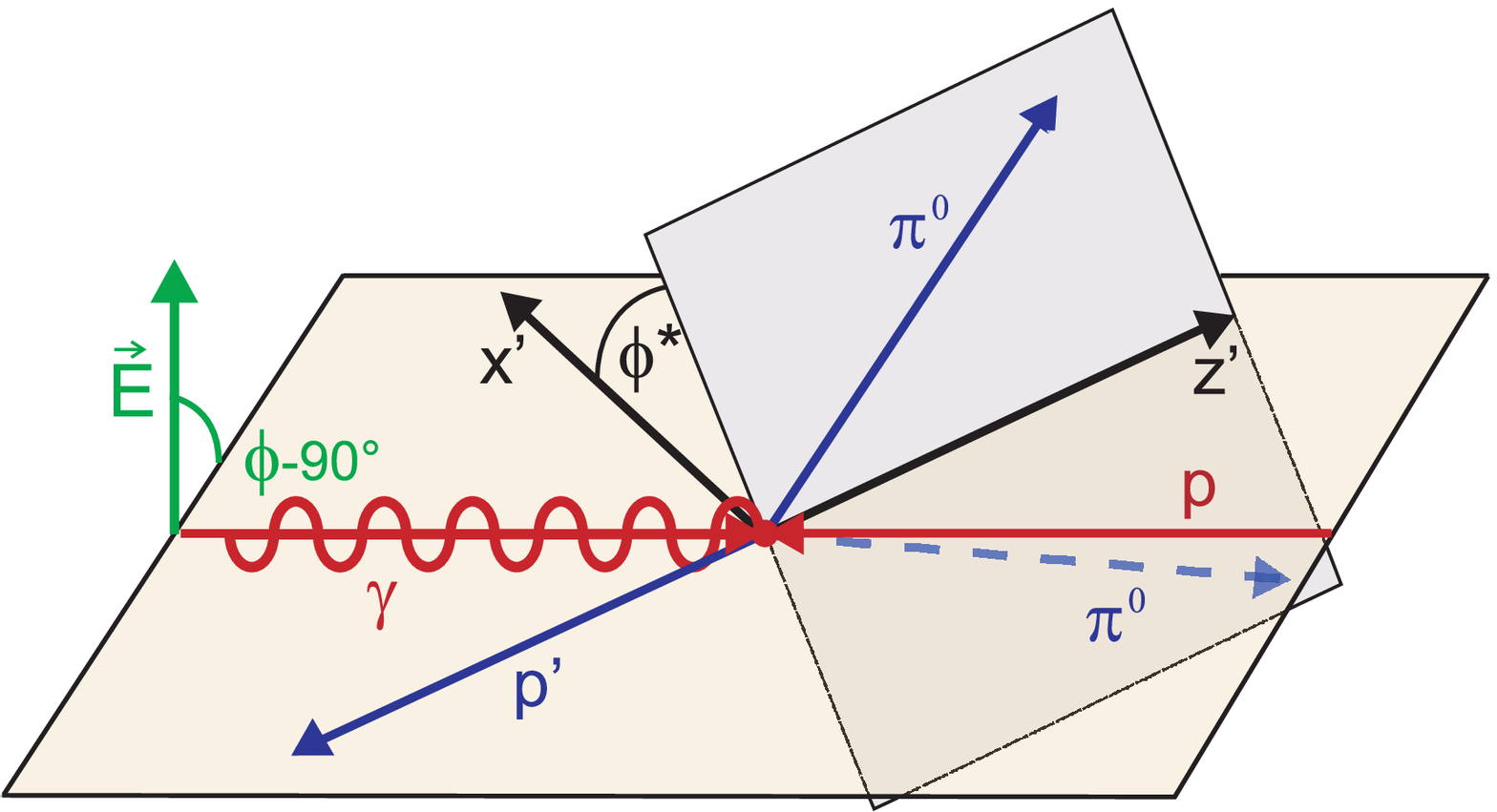}

 \begin{small}\put(87, 42){\bf \sffamily (a)}\end{small}
\end{overpic}
      \begin{overpic}[width= 8 cm,
]
{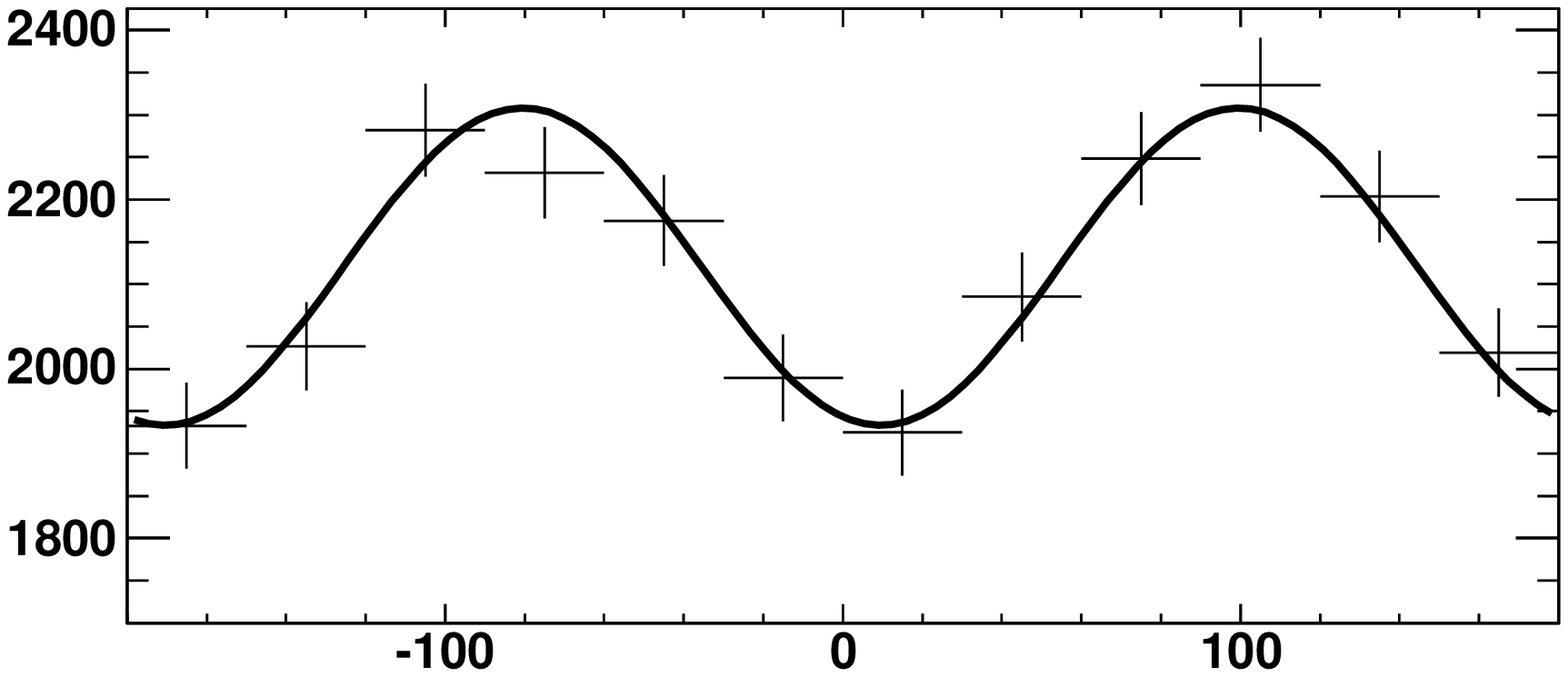}
\begin{small}\put(84, 42){\bf \sffamily (b)}\end{small}
\begin{small}\put(81.5, -1){\bf \sffamily \boldmath$\phi [^{\circ}]$\unboldmath}\end{small}

%

\end{overpic}

\caption{(a) Kinematics of the three-body final state \cite{Gutz-Is} \cite{Helicity},
 the angle $\phi^{*}$, only occurring for a three body final state is the angle between the production
  plane formed by the incoming photon and the recoiling particle (proton in this case) and the decay plane formed 
  by the particles in the final state. (b) Example of a $\phi$ distribution in the interval 
  $144^{\circ} < \phi^{*} < 162^{\circ}$ and $E_{\gamma} = 970-1200 \rm \, MeV$.}
\label{fig: kinematics}
\end{figure}

For the understanding of the excitation spectrum and the properties of baryons photoproduction
reactions play an important role. In comparison to single meson photoproduction
the importance of the double-meson photoproduction increases at higher energies due to higher cross-sections. 
The investigation of the double-meson final states also allows access to cascading decays of resonances
via intermediate states such as the $\Delta(1232)$ and $D_{13}(1520)$. According to \cite{RobOed-PRC} for a complete experiment 
in the two-meson photoproduction 15 independent observables have to be measured. 
The polarization observables $I^{s}$ and $I^{c}$, accessible for two-meson final states in the case of linearly
polarized photon beam and unpolarized target, were measured with the CBELSA/TAPS experiment for the reaction $\vec{\gamma}\mathrm{p}\rightarrow\mathrm{p}\pi^{0}\eta$
\cite{Gutz-Is}. Data on the reaction $\vec{\gamma}\mathrm{p}\rightarrow\mathrm{p}\pi^{0}\pi^{0}$
with linearly polarized photons and unpolarized target were obtained by the GRAAL experiment, where the observable $\Sigma$ 
was measured \cite{Assafiri}. As discussed in \cite{Gutz-Is}, the observable 
$I^{c}$ reduces to $\Sigma$ in a quasi two body consideration. The cross section for the case of a linearly polarized photon beam 
and an unpolarized target \cite{RobOed-PRC} can be written as:
\begin{equation}
 \frac{\mathrm{d}\sigma}{\mathrm{d}\Omega} = \left(\frac{\mathrm{d}\sigma}{\mathrm{d}\Omega}\right)_{0}\left(1+\delta_{l}\left(I^{c}\cos2\phi+I^{s}\sin2\phi\right)\right),
\label{eq: cs three body}
\end{equation}
where $\left(\frac{\mathrm{d}\sigma}{\mathrm{d}\Omega}\right)_{0}$ is the unpolarized cross-section, $\delta_{l}$
is the degree of linear polarization of the incoming photon. In comparison to the single meson photoproduction where the incoming photon and outgoing
particles are in the same plane, in the double meson photoproduction an additional plane occurs. Consideration of the $\phi$ distributions in the different intervals of 
angle $\phi^{*}$, being the angle between production and decay planes (see Figure \ref{fig: kinematics}a) gives access to the observables 
$I^{s}$ and $I^{c}$ as function of the angle $\phi^{*}$. An example of a $\phi$ distribution in a limited $\phi^{*}$
range, illustrating the contributions of sine and cosine terms, is shown in Figure \ref{fig: kinematics}b.

\section{Experimental Setup}

The CBELSA/TAPS experiment was performed at the electron accelerator ELSA in Bonn \cite{Hillert-EPJA}.
Linearly polarized photons were produced via coherent bremsstrahlung on a
diamond crystal. The energy of these photons was determined by a scintillator based tagging system. 
The tagged photons were impinging on a 5 cm liquid hydrogen target \cite{Kopf-PhD} surrounded by the
Crystal Barrel (CB) \cite{Aker-NIM} and TAPS \cite{TAPS} calorimeters. 
The CB consisting of 1290 CsI(Tl) crystals covers the polar angular range $ 30^{\circ} < \theta < 168^{\circ}$,
TAPS with 528 $\rm BaF_{2}$ crystals covers $5^{\circ} < \theta < 30^{\circ}$.
For charge identification the target is surrounded by a 3 layer scintillating fiber detector \cite{Suft-NIM} 
and each of the TAPS crystals is equipped 
with a plastic scintillator in front of it. In this analysis data with two linear polarization settings
were considered reaching a maximum polarization of 49.2\% at an energy of 1300 MeV and of 38.7 \% at 1600 MeV. 

\section{Data analysis}
The reaction $\gamma \mathrm{p} \rightarrow \mathrm{p} \pi^{0} \pi^0$ has been selected asking for events with
four or five clusters in the calorimeters. 
The events were retained if two of the two particle invariant masses resulted in the pion mass
within $\pm \rm 35\,MeV$ and the missing mass of the fifth particle was consistent with 
the proton mass within $\pm \rm 100\,MeV$. In case of five cluster events the missing proton direction had to  
match with the additional hit in one of the calorimeters. 
For four cluster events with existing inner detector hit the missing proton direction had to match with this hit.
To further reduce the background for four cluster events additional cuts derived from Monte Carlo simulations taking 
advantage of the correlation between the momentum and polar angle of the missing proton  were performed \cite{Sokhoyan-PhD}. 
After preselection, the data was subjected to a kinematic fit \cite{Pee-EPJA} assuming that the interaction took place in the target center. 
The events that exceeded a Confidence Level (CL) of 10\% for the $\gamma \mathrm{p} \rightarrow \mathrm{p} \pi^{0} \pi^0$ hypothesis
were retained as long as the CL of $\gamma \mathrm{p} \rightarrow \mathrm{p} \pi^{0} \pi^0$ did as well
exceed the CL for the competing $\gamma \mathrm{p} \rightarrow \mathrm{p} \pi^{0} \eta$ hypothesis.
The proton direction resulting from the fit had to agree with the direction of the proton reconstructed 
in one of the calorimeters where available. Additionally, cuts on the energy deposited in the calorimeters 
by the particle selected to be a proton by kinematic fit and on the number of crystals in 
the clusters produced by protons, were derived from Monte Carlo simulations. 
The background level was determined to be well below 1\%. For the determination of polarization observables 
the cross-section (see eq. \ref{eq: cs three body}) was parametrized as:
 \begin{equation} 
 f(\phi) = A + P(B\cos2\phi + C\sin2\phi),
 \label{eq: cs three body param}        
 \end{equation}
where P is the degree of the linear polarization. 

\section{Results}

\begin{figure}[hbtp]
      \begin{overpic}[width=\textwidth,
]
{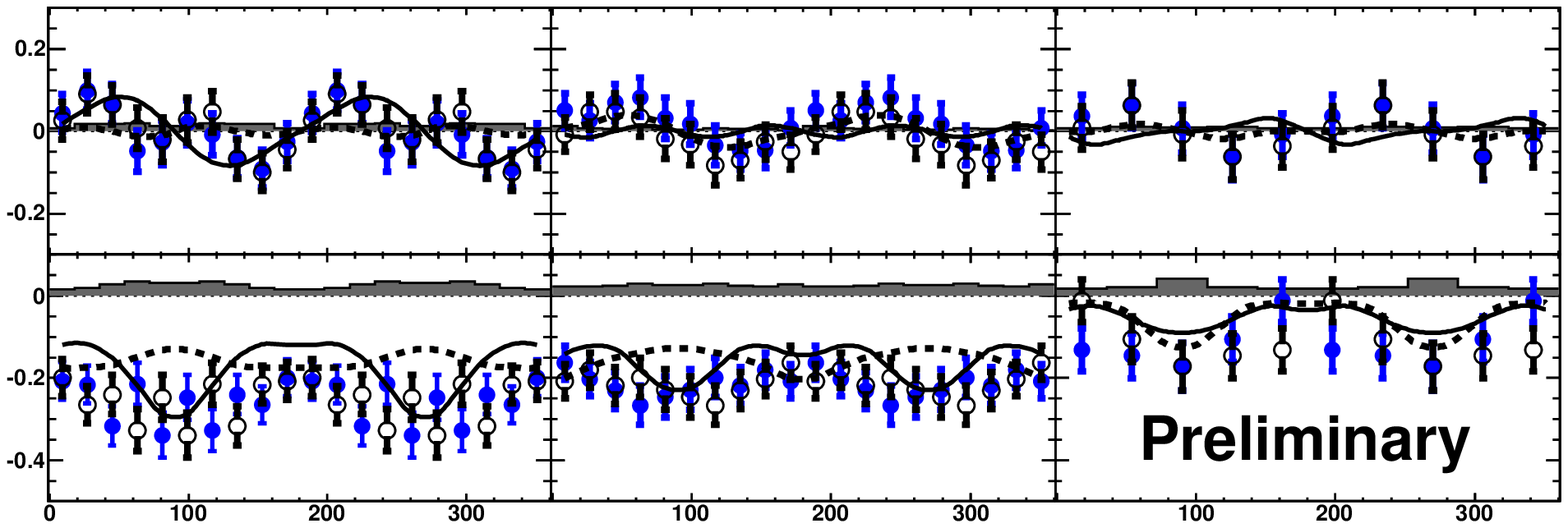}

%
%

\begin{small}\put(91.9, -1.2){\bf \sffamily \boldmath$\phi^{*} [^{\circ}]$\unboldmath}\end{small}

\begin{small}\put(-1, 14){\bf \sffamily \boldmath$I^{c}_{p}$\unboldmath}\end{small}
\begin{small}\put(-1, 29){\bf \sffamily \boldmath$I^{s}_{p}$\unboldmath}\end{small}

\begin{small}\put(7,33.5){\bf \sffamily \boldmath$\mathsf{E_{\gamma} = 970-1200}\rm \, MeV$ \unboldmath}\end{small}
\begin{small}\put(38,33.5){\bf \sffamily \boldmath$\mathsf{E_{\gamma} = 1200-1450}\rm \, MeV$ \unboldmath}\end{small}
\begin{small}\put(69,33.5){\bf \sffamily \boldmath$\mathsf{E_{\gamma} = 1450-1650}\rm \, MeV$ \unboldmath}\end{small}

%

\end{overpic}
\caption {Preliminary results on the polarization observables  $I^{s}$ (top) and $I^{c}$ (bottom), 
with the proton considered to be the recoiling particle. 
Solid symbols: extracted directly from the data, open symbols: derived from  
$I^{c}(\Phi^{*}) = I^{c}(2\pi - \Phi^{*})$ and $I^{s}(\Phi^{*}) = -I^{s}(2\pi - \Phi^{*})$.
The bars show the systematic errors giving an estimate of the effect uncovered phase space areas may have on the result. 
Solid curve: Bonn-Gatchina PWA solution with the dominant $D_{33}(1700)\rightarrow \Delta \pi$ D-wave decay, 
  dashed curve: $D_{33}(1700)\rightarrow \Delta \pi$ S-wave decay.}
\label{fig: Is and Ic proton}
\end{figure}

\begin{figure}[hbtp]
      \begin{overpic}[width=\textwidth,
]
{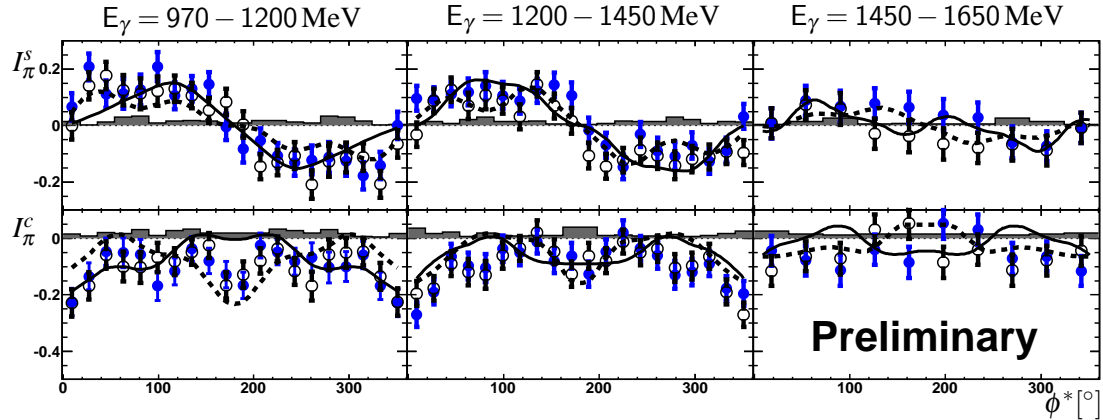}
%
%

\begin{small}\put(91.9, -1.2){\bf \sffamily \boldmath$\phi^{*} [^{\circ}]$\unboldmath}\end{small}
\begin{small}\put(-1, 14.5){\bf \sffamily \boldmath$I^{c}_{\pi}$\unboldmath}\end{small}
\begin{small}\put(-1, 29.5){\bf \sffamily \boldmath$I^{s}_{\pi}$\unboldmath}\end{small}

\begin{small}\put(7,33.5){\bf \sffamily \boldmath$\mathsf{E_{\gamma} = 970-1200}\rm \,MeV$ \unboldmath}\end{small}
\begin{small}\put(38,33.7){\bf \sffamily \boldmath$\mathsf{E_{\gamma} = 1200-1450}\rm \,MeV$ \unboldmath}\end{small}
\begin{small}\put(69,33.7){\bf \sffamily \boldmath$\mathsf{E_{\gamma} = 1450-1650}\rm \,MeV$ \unboldmath}\end{small}

%

\end{overpic}
\caption{Preliminary results on the polarization observables  $I^{s}$ (top) and $I^{c}$ (bottom), with the pion considered to be the recoiling
particle, notation as in Figure \ref{fig: Is and Ic proton}.}
\label{fig: Is and Ic pi}
\end{figure}

Figure \ref{fig: Is and Ic proton} shows the observables $I^{s}$ and $I^{c}$ as function of the angle $\phi^{*}$
(see Figure \ref{fig: kinematics}a) for the proton recoiling case. Solid points are extracted from the data directly, the open points from 
the conditions $I^{s}(\phi^{*}) = -I^{s}(2\pi - \phi^{*})$ and $I^{c}(\phi^{*}) = I^{c}(2\pi - \phi^{*})$
which result from symmetry and parity constraints \cite{Gutz-Is}. Good agreement between the two 
sets indicates absence of strong systematic effects in the data. Also, due to the indistinguishability
of $\pi^{0}$s, in the proton recoiling case, additional symmetry
conditions $I^{s}(\phi^{*}) = I^{s}(\phi^{*} + \pi)$ and  $I^{c}(\phi^{*}) = I^{c}(\phi^{*} + \pi)$ occur.
In Figure \ref{fig: Is and Ic pi} the data for the pion recoiling case are shown.
In the studies of the unpolarized data  with Bonn-Gatchina Partial-Wave Analysis (PWA) \cite{Anisovich} \cite{Thoma-2pi}  it was found that the $D_{13}(1520)$ decays into 
$\Delta \pi$ in S- and D-waves with similar strength. Also, for the $D_{13}(1700)$ decays into $\Delta \pi$ the 
contribution of D-wave was observed to be stronger than the contribution of S-wave. 
These results contradict naive expectations expecting D-wave decay modes to be suppressed 
due to the orbital angular momentum barrier. 
To further investigate the contributions of S- and D-waves in the $D_{33}(1700)$ decay,
which were not determined in \cite{Thoma-2pi} due to ambiguous solutions in the PWA,
the data on $I^{s}$ and $I^{c}$ were compared to PWA solutions with D-wave dominance (solid) and S-wave dominance (dashed).
These predictions were produced without including these data in the PWA.
Noticeable differences are observed between the two predictions, indicating sensitivity of the PWA
to the observables $I^{s}$ and $I^{c}$. From this comparison one can also conclude that
none of the solutions is clearly favored, but that the data provides valuable input for the PWA.

\section{Summary and conclusions}
Polarization observables $I^{s}$ and $I^{c}$ were measured for the first time in the reaction 
$\vec{\gamma} \mathrm{p} \rightarrow \mathrm{p} \pi^{0} \pi^0$ with the CBELSA/TAPS experiment. 
A comparison of $I^{s}$ and $I^{c}$ with predictions of the Bonn-Gatchina PWA showed
that the data will provide new constraints for the PWA. This work was supported by the
 Deutsche Forschungsgemeinschaft (DFG) within SFB/TR16.

\bibliographystyle{aipproc}   

\begin{thebibliography}{9}
 
 \bibitem{RobOed-PRC}
W.~Roberts, T.~Oed,
Phys.\ Rev.\  C {\bf 71}, 055201 (2005).
 
\bibitem{Gutz-Is}
E.~Gutz, V.~Sokhoyan, H.~van Pee {\it et al.},
Phys.\ Lett.\ B.\  {\bf 687}, 11 (2010).

\bibitem{Assafiri}
  Y.~Assafiri  {\it et al.},
  Phys.\ Rev.\ Lett.\  {\bf 90}, 222001 (2003).

\bibitem{Helicity}
  S.~Strauch {\it et al.},
  Phys.\ Rev.\ Lett. {\bf 95}, 162003 (2005);
D.~Krambrich {\it et al.},
  {\em ibid.}. {\bf 103}, 052002 (2009).
  
\bibitem{Hillert-EPJA}
W.~Hillert,
Eur. Phys. J. A \textbf{28}, 139 (2006).

\bibitem{Kopf-PhD}
B.~Kopf,
Ph.D. thesis, Dresden (2002).

\bibitem{Aker-NIM}
E.~Aker {\it et al.},
Nucl. Inst. and Meth. A  \textbf{321}, 69 (1992).

\bibitem{TAPS}
R.~Novotny,
IEEE Trans. Nucl. Sci. \textbf{NS-38}, 379 (1991);
A.R.~Gabler {\it et al.},
 Nucl. Inst. and Meth. A \textbf{346}, 168 (1994).

\bibitem{Suft-NIM}
G.~Suft {\it et al.},
Nucl. Inst. and Meth. A \textbf{538}, 416 (2005).

\bibitem{Sokhoyan-PhD}
V.~Sokhoyan,
Ph.D. thesis in preparation, Bonn (2011).

\bibitem{Pee-EPJA}
H.~van Pee {\it et al.},
Eur. Phys. J. A \textbf{31}, 61 (2007).


\bibitem{Anisovich}
 A.V.~Anisovich {\it et al.},
 Eur. Phys. J. A \textbf{24}, 111 (2005);
 Eur. Phys. J. A \textbf{34}, 243 (2007).

\bibitem{Thoma-2pi}
U.~Thoma {\it et al.},
Phys.\ Lett.\ B.\  {\bf 659}, 87 (2008).


\end{thebibliography}

\end{document}